\documentclass[twocolumn,showpacs,prb,superscriptaddress]{revtex4}
\usepackage{subfigure}
\usepackage{amsmath}
\usepackage{graphicx}
\usepackage{rotating}

\begin{document}

\title{Tuning impurity states in bilayer graphene.}

\author{Hari P. Dahal}
\affiliation{Department of Physics, Boston College, Chestnut Hill,
 MA, 02467}
 \affiliation{Theoretical Division, Los Alamos National Laboratory,
Los Alamos, New Maxico 87545}

\author{A. V.  Balatsky}
\affiliation{Theoretical Division, Los Alamos National Laboratory,
Los Alamos, New Maxico 87545} \affiliation{Center for Integrated
Nanotechnology, Los Alamos National Laboratory, Los Alamos, New
Maxico 87545}

\author{Jian-Xin Zhu}
\email{jxzhu@lanl.gov} \homepage{http://theory.lanl.gov}
 \affiliation{Theoretical Division, Los
Alamos National Laboratory, Los Alamos, New Maxico 87545}

\begin{abstract}
We study the impurity states in bilayer graphene in the unitary
limit using Green's function method. Unlike in single layer
graphene, the presence of impurities at two non-equivalent sites
in bilayer graphene produce different impurity states which is
understood as the change in the band structure due to interlayer
hopping of electrons. The impurity states can also be tuned by
changing the band structure of bilayer grahene through external
electric field bias.
\end{abstract}

\maketitle

\section{Introduction}

In recent years, the fabrication of few layers of graphene systems
\cite{novoselov2005,novoselov22005,novoselov12006} has attracted a
lot of attention to study the electronic properties of these
systems. The electrons in single graphene show some unconventional
electronic properties such as the half integer quantum Hall
effect, and Klein paradox. The electrons in bilayer systems also
show some interesting properties as seen in quantum Hall effect.
Many properties of single and bilayer graphenes also differ
because of the different crystal structure. In this communication
we present a systematic study of impurity effects in bilayer
graphene and contrast it with those of the single layer
counterpart.

\begin {figure}
\includegraphics[width= 6.0 cm, height = 4.50cm, angle=0]{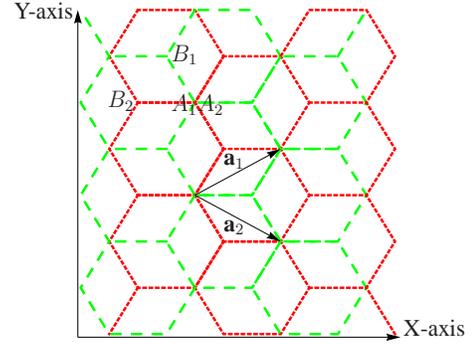}
\caption{(Color online) Lattice structure of bilayer graphene
under consideration for a tight binding calculation. The green
(dashed) line forms the top layer and the red (dotted) line forms
the bottom layer.} \label{FIG:lattice}
\end{figure}

The fundamental difference between single and bilayer graphene
originates from their crystal structure. Single layer graphene is
an atomically thin two-dimensional hexagonal packing of $sp^2$
bonded carbon atoms. It is the building block of (multi-layer)
graphene. One unit cell of single layer graphene has two
non-equivalent lattice sites (A, B). As s result, the electron
wave function is spinor-like, where the sublattice index plays the
role of pseudo-spin. The tight binding calculation
\cite{wallace1947} shows that the electrons in single layer
graphene disperse linearly, i.e. ($E_k=\pm v_F k $), where
$v_F=\frac{3ta}{2}=5.8$ $eV \AA$ is the Fermi velocity, $t =3.0
eV$ is the nearest neighbor hopping energy in the plane, and $a$
is the lattice constant, and hence are called the massless Dirac
fermions. The bilayer graphene, as shown schematically in Fig.
\ref{FIG:lattice}, can be thought of as a stacking of two
identical single layer graphenes in the third dimension. In one of
the common ways of layer stacking, known as Bernal stacking, only
one of the non-equivalent lattice sites (site-A) stay on top of
each other, another site (B) lies in the middle of the hexagon of
the other layer.\cite{nilsson2006c} The electron can hop between
the layers along the bonding of these two A-sties with a hopping
energy ($t_\bot$), which is about ten times smaller than the
hopping energy along the plane. This interlayer hopping hybridizes
the $p_z$ orbital of the carbon atom at site-A resulting in
different dispersion relation of the electrons, ($E_k = \pm
\frac{t_\bot}{2} \pm \sqrt{ \frac{t_\bot ^2}{4} + v_F^2 k^2}$).
\cite{nilsson2006b, mccann2006a, mccann2006b} Two of the branches
of the electronic band touch each other at the Fermi energy,
whereas the other two branches become gapped with energy gap equal
to $t_\bot$. The dispersion relation of the electrons
corresponding to the gapless branches can be expressed in
parabolic form at low momenta. The electron energy in this case
have parabolic dispersion at low momenta. In addition to this
difference of the band structure between single and bilayer
graphenes, the bilayer system gives a freedom of tailoring the
band structure by applying an external electric field bias (V) on
the two layers. The dispersion relation in the presence of the
external field bias becomes, \cite{mccann2006b, nilsson2007a}
\begin{equation}
E_K=\pm \sqrt{\frac{V^2}{4}+\frac{t_\bot^2}{2}+v_F^2 k^2\pm \sqrt{
\frac{t_\bot^4}{4}+(V^2+t_\bot^2)v_F^2 k^2}},
\label{Eq:approxeigenvalues}
\end{equation}
which shows that the two valence (conduction) bands shift below
(above) the Fermi energy by $V/2$.

To understand the electronic property of these systems, it is
important to study the revealing impurity effects. The impurity
states in single layer graphene has been studied by Wehling
\textit{et al.}\cite{wehling2007} and by Bena.\cite{bena2007} It
has been shown that if an impurity is introduced on a lattice
site, a virtually bound impurity resonance state can be induced,
which can be seen as an enhancement of the local density of states
(LDOS) at the neighboring site. This effect is symmetric with
respect to the sublattice the impurity is on, i.e., an impurity at
site B shows exactly the same effect as it is at site A. In
bilayer graphene, due to the bonding between the sites A of the
two layers, the site A and B in each layer are no longer
equivalent. The interlayer hopping distinctly differentiates the
two sites. So we expect to see different impurity states with
respect to the location of the impurity. In addition, it is
natural to expect that the external bias which changes the band
structure also modifies the properties of the impurity states.
This motivates us to study the impurity states in bilayer graphene
as a function of both $t_\bot$ and $V$. The impurity effect in
single layer graphene is contained in our discussion as a special
case of bilayer graphene where $t_\bot =0$.

The outline of the paper is as follows: In Sec. II, we present a
Green's function method to study the impurity effects. For the
clean system, the results from the Green's function method is
benchmarked with the exact diagonalization. In Sec. III, we
present a systematic set of results and discussions. Concluding
remarks are given in Sec. IV.

\section{Theoretical method}

In this section we describe the theoretical method used to
calculated the LDOS in the bilayer graphene using Green's
function. We need to find the form of the Hamiltonian to define
the free particle Green's function. The Green's function in the
presence of the impurity potential is obtained by using the
Dyson's equation. We derive a tight binding Hamiltonian using the
lattice structure of the bilayer graphene as shown in Fig.
\ref{FIG:lattice}. The basis vectors are chosen to be,
$a_1=(\frac{3a}{2},\frac{-\sqrt{3}a}{2})$, and
$a_1=(\frac{3a}{2},\frac{\sqrt{3}a}{2})$. The correspond basis
vectors of the reciprocal space is given by
$b_1=(\frac{2\pi}{3a},\frac{-2\pi}{\sqrt{3}a})$, and
$b_2=(\frac{2\pi}{3a},\frac{2\pi}{\sqrt{3}a})$.  Sites $A_1$ and
$A_2$ are connected along the $\textit{z}$-direction, the
electrons have $t_\bot$ hopping energy in this direction. We have
assumed that the layer which has lattice sites $A_1$, $B_1$ is
biased with $V/2$ and the layer which has lattice sites $A_2$,
$B_2$ with $-V/2$ so that the potential difference between the two
layers is $V$. For the convenience of discussion, these two layers
are called the top and bottom layers, respectively.

The tight binding equations for this lattice structures in the
clean case can be written as,
\begin{subequations}
\begin{widetext}
\begin{equation}
\frac{V}{2}\phi_{A_{1}} -
t[2\exp(\frac{-ik_xa}{2})\cos(\frac{k_ya\sqrt{3}}{2}) +
\exp(ik_xa)]\phi_{B_{1}} - t_\bot \phi_{A_{2}}   = E \phi_{A_{1}},
\end{equation}

\begin{equation}
 - t[2\exp(\frac{ik_xa}{2})\cos(\frac{k_ya\sqrt{3}}{2})
+ \exp(-ik_xa)]\phi_{A_{1}} + \frac{V}{2}\phi_{B_{1}} = E
\phi_{B_{1}},
\end{equation}

\begin{equation}
 - t_\bot \phi_{A_{1}} +  \frac{-V}{2}\phi_{A_{2}}
 - t[2\exp(\frac{ik_xa}{2})\cos(\frac{k_ya\sqrt{3}}{2})
+ \exp(-ik_xa)]\phi_{B_{2}} = E \phi_{A_{2}},
\end{equation}

\begin{equation}
 - t[2\exp(\frac{-ik_xa}{2})\cos(\frac{k_ya\sqrt{3}}{2}) +
\exp(ik_xa)]\phi_{A_{2}} + (\frac{-V}{2})\phi_{B_{2}} = E
\phi_{B_{2}}.
\end{equation}
\end{widetext}
\end{subequations}
These equations can be written in a matrix form,
\begin{equation}
 \widehat{H}_{\textbf{k}}\left(%
\begin{array}{c}
  \phi_{A_{1}} \\
  \phi_{B_{1}} \\
  \phi_{A_{2}} \\
 \phi_{B_{2}} \\
\end{array}%
\right) = \varepsilon_n \left(%
\begin{array}{c}
  \phi_{A_{1}} \\
  \phi_{B_{1}} \\
  \phi_{A_{2}} \\
  \phi_{B_{2}} \\
\end{array}%
\right).
\end{equation}
Here,
\begin{equation}
\widehat{H}_{\textbf{k}}=\left(%
\begin{array}{cccc}
  V/2 &\widetilde{t} & -t_\bot & 0 \\
  \widetilde{t}^* & V/2 & 0 & 0 \\
  -t_\bot & 0 & -V/2 & \widetilde{t}^* \\
  0 & 0 & \widetilde{t} & -V/2 \\
\end{array}%
\right),
 \label{Eq:hamiltonian}
\end{equation}
with $\widetilde{t} = -
t[2\exp(\frac{-ik_xa}{2})\cos(\frac{k_ya\sqrt{3}}{2}) +
\exp(ik_xa)]$ and $\widetilde{t}^*$ is the complex conjugate of
$\widetilde{t}$.

The Green's function corresponding to this Hamiltonian can be
expressed as,
\begin{equation}
\widehat{G}^{(0)}(\textbf{k},\omega)= [\omega
\widehat{1}-\widehat{H}_{\textbf{k}}]^{-1},
\end{equation}
where $\widehat{1}$ represents the unit matrix. As long as all
introduced impurities (up to 4) are located within a single cell,
we can express the local Green's function exactly through the
T-matrix method, which leads to
\begin{equation}
\widehat{G}_{ij}(\omega)=\widehat{G}_{ij}^{(0)}(\omega)+
\widehat{G}_{i0}^{(0)}(\omega)\widehat{T}(\omega)G_{0j}^{0}(\omega).
\end{equation}
Here the T-matrix
$\widehat{T}(\omega)=\widehat{U}(1-G^{(0)}(\omega)\widehat{U})^{-1}$,
the local Green's function
$\widehat{G}_{ij}^{(0)}(\omega)=\frac{1}{N}\sum_{\textbf{k}}\widehat{G}^{(0)}(\textbf{k},\omega)$
where the summation is over the first Brillouin zone, and
$\widehat{U}$ is the matrix representation of the impurity
potential. The local density of states at different sites is given
by, $ N_{A1}= \frac{-1}{\pi} G_{11}(\omega+i\gamma)$,
$N_{B1}=\frac{-1}{\pi} G_{22}(\omega+i\gamma)$,
$N_{A2}=\frac{-1}{\pi} G_{33}(\omega+i\gamma)$,
$N_{B2}=\frac{-1}{\pi} G_{44}(\omega+i\gamma)$, where $\gamma$ is
the lifetime broadening. For numerical calculation all relevant
energies are measured in terms of the intralayer hopping energy of
the electron, t. For simplicity we use the impurity potential
close to the unitary limit, $U=100$. The number of $1024\times
1024$ $\textbf{k}$ points are used in the Brillouin zone. The
intrinsic lifetime broadening of $\gamma = 0.005$ is taken. When
we present results, unless otherwise stated, the panel $a$ to $d$
represent the LDOS at sites $A_1$, $B_1$, $A_2$, and $B_2$,
respectively. In each figure, the DOS is plotted (along the
vertical axis) as a function of energy $E$ (along the horizontal
axis).

We can also calculate DOS in the absence of impurities using
eigenvalues and eigenvectors of Hamiltonian matrix, Eq.
(\ref{Eq:hamiltonian}). A little algebra yields the eigenvalues
\begin{equation}
\varepsilon_n = \mp
\sqrt{\frac{V^2}{4}+\frac{t_\bot^2}{2}+\widetilde{t}
\widetilde{t}^* \mp
\sqrt{\frac{t_\bot^4}{4}+(V^2+t_\bot^2)\widetilde{t}
\widetilde{t}^*}}, \label{Eq:eigenvalues}
\end{equation}
where $\varepsilon_n$ represents the four eigenvalues. Note that
the linearization of Eq. (\ref{Eq:eigenvalues}) near the corner of
the Brillouin zone reduces to Eq. (\ref{Eq:approxeigenvalues}).

The corresponding eigenvectors are given by a general equation,
\begin{equation}
\left(%
\begin{array}{c}
  \phi_{A1, \varepsilon_n}  \\
  \phi_{B1, \varepsilon_n}  \\
  \phi_{A2, \varepsilon_n}  \\
  \phi_{B2, \varepsilon_n}  \\
\end{array}%
\right) = \left(%
\begin{array}{c}
  |\widetilde{t}|^2-(\frac{V}{2}+\varepsilon_n)^2 \\
  |\widetilde{t}|^2(\varepsilon_n-\frac{V}{2})+(\frac{V}{2}+\varepsilon_n)(t_\bot^2+\frac{V^2}{4}-{\varepsilon_n}^2) \\
  \frac{\frac{V}{2}+\varepsilon_n}{\widetilde{t}} \\
  1 \\
\end{array}%
\right)
\end{equation}
where substituting four different values of $\varepsilon_n$ gives
the four sets of eigenvectors. We normalize thus obtained $4
\times 4$ matrix of the eigenvectors and calculate the density of
states on four nonequivalent sites using the standard equation,
\begin{equation}
N_{\alpha \beta}(E)= \frac{1}{N \pi} \sum_{\textbf{k},
\varepsilon_n}|\phi_{\alpha \beta, \varepsilon_n}(\textbf{k})|^2
\biggl{[}\frac{\gamma}{(E-\varepsilon_n)^2 + \gamma ^2}\biggr{]},
\end{equation}
where $\alpha = A, B$, $\beta = 1, 2$, $\phi_{\alpha \beta,
\varepsilon_n}(\textbf{k})$ is the normalized eigen-vectors. We
used this DOS to check the result obtained by Green's functions
method in the absence of impurity.

\section{Results and Discussions}
\subsection{ Clean limit}

\begin {figure}
\includegraphics[width= 8.50 cm, height = 6.50cm]{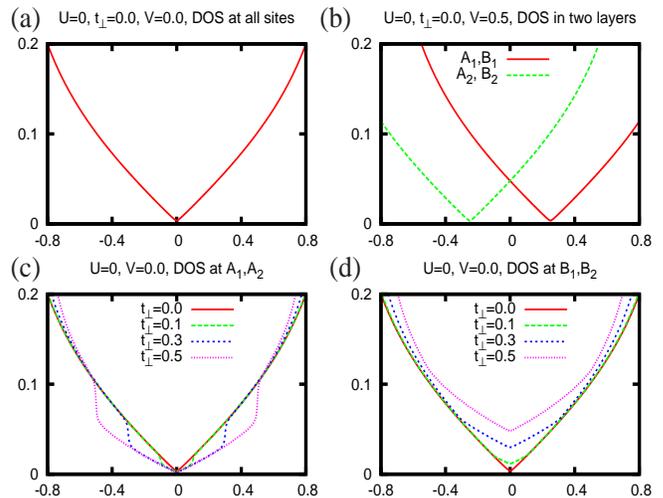}
\caption{(Color online) The density of states as a function of
energy is shown. a) For $t_\bot =0$ the DOS at all sites is equal,
and vanishes linearly at the Fermi energy. b) For $t_\perp =0$,
and $V \neq 0$, the DOS curve for sites of the top (bottom) layer
shifts above (below) the Fermi energy by $V/2$ creating finite DOS
at the Fermi level. c) For $t_\bot\neq 0$ DOS at $A$ sites of both
layers is equal and for $-t_\perp<E<t_\perp$ it decreases compared
to the single layer case. d) DOS at sites $B$ of both layers is
equal and for $-t_\perp<E<t_\perp$ it increases compared to the
single layer case.}
 \label{FIG:u=0LDOSsingleandbilayer}
\end{figure}

Before getting in to the discussion of the impurity states, we
first discuss the effects of $V$ and $t_\bot$ on the DOS in the
absence of the impurity. For $V=0$, and $t_\bot=0$, the DOS at
every sites is equal and it vanishes linearly in $E$ close to the
Fermi energy as shown in Fig.
\ref{FIG:u=0LDOSsingleandbilayer}(a). For $t_\bot=0$ but $V \neq
0$ the DOS in two layers is different. The overall variation of
the DOS is still preserved but the DOS curve of upper (lower)
layer shifts towards positive (negative) energy region by $V/2$,
Fig. \ref{FIG:u=0LDOSsingleandbilayer}(b). So, the minima in the
DOS for upper (lower) layer lies at $E=|V|/2$. This results in a
finite density of states at the Fermi energy. The shift in the
position of the DOS minima will have nontrivial influence on
impurity states as discussed below.

\begin {figure}
\includegraphics[width= 8.5 cm, height = 10.0cm]{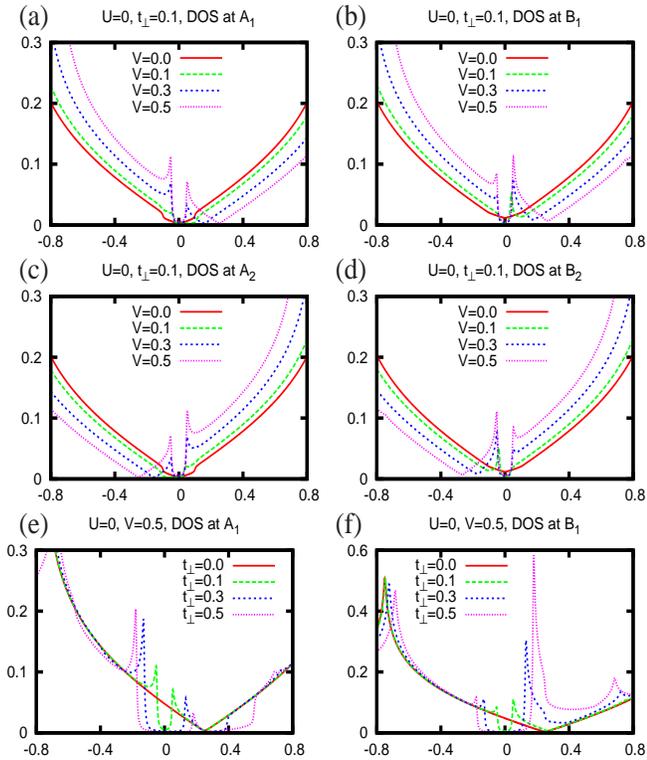}
\caption{(Color online) LDOS in the absence of impurity at fixed
$t_\bot =0.1$ and different $V$. The minimum of the DOS is at
$|V|/2$. Small enhancement of the DOS occurs close to Fermi energy
with the increase in $V$. We will show later that the position of
this enhancement depends on $t_\bot$. There is a rearrangement of
DOS between A, B sites. e)and f) LDOS at site $A_1$ and $B_1$
(respectively) at fixed $V=0.5$ and different $t_\bot$ is shown.
We see gap opening around the Fermi energy. The magnitude of the
gap increases with the increase in $t_\bot$. LDOS at site $A_2,
B_2$ can be obtained by mirror inversion to the curves in Fig.
\ref{FIG:U=0difftdiffv}(e) and \ref{FIG:U=0difftdiffv}(f) about
the axis of $E=0$, respectively.} \label{FIG:U=0difftdiffv}
\end{figure}

The difference in the DOS of single and bilayer graphene can be
studied by using finite value of $t_\bot$. In Fig.
\ref{FIG:u=0LDOSsingleandbilayer}(c-d) we show the DOS at four
sites for various values of $t_\bot$ but fixed $V=0$. The DOS at
$A_1, A_2$ is equal and similarly DOS at $B_1, B_2$ is equal but
those at $A$ sites and $B$ sites are not equal in the energy range
$-t_\bot < E < t_\bot$. In this energy range, the DOS at $A$ sites
is smaller than that at $B$ sites. In particular, at the Fermi
energy, the DOS at $B$ sites is finite but that at $A$ sites is
zero. These effects can be understood in terms of the difference
in the band structure caused by finite $t_\bot$. The band
structure corresponding to sites $A_1$ and $A_2$ describe the
anti-bonding states characterized by band gaps of $\pm t_\perp$
which results in the decreased density of states compared to that
of the single layer graphene where corresponding bands are
gapless. On the other hand, the band structure corresponding to
the sites $B_1$ and $B_2$ is gapless at the Fermi energy and has
more flat band compared to the single layer case resulting in
finite density of states at the Fermi level. An interesting point
is that the DOS at $B$ sites is also linear even though the
dispersion relation of the electron corresponding to $B$ sites can
be approximated by parabolic dispersion at low momenta.

Here, we also discuss the combined effect of bias and interlayer
hoping on the DOS. In Fig. \ref{FIG:U=0difftdiffv}(a-d) we have
shown the DOS at four sites for fixed $t_\bot=0.1$ and different
$V$. In Fig. \ref{FIG:U=0difftdiffv}(e-f) we have also shown the
DOS at $A_1$ and $B_1$ for fixed $V=0.5$ and different $t_\bot$.
Some new features are seen in these figures. The DOS at all sites
gets modified compared to the unbiased case (compare with Fig.
\ref{FIG:u=0LDOSsingleandbilayer}(c) and (d)).  There is a shift
of DOS minimum to $E=\pm V/2$ compared to $V=0$ case. There is
also a gap opening up around the Fermi energy which increases with
the increase in $t_\bot$. These unique features of band structure
will lead to non-trivial impurity states.

\begin {figure}
\includegraphics[width= 8.5 cm, height = 6.50cm]{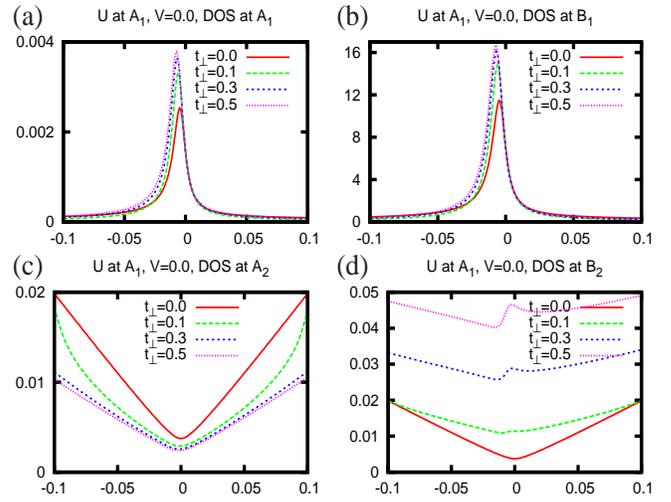}
\caption{(Color online) For single impurity at site $A_1$ DOS  as
a function of energy is shown. DOS at the impurity site, $A_1$, is
very small. DOS at the neighboring site in the same layer, $B_1$,
is very high which is the signature of the impurity resonance. The
resonance is close to the Fermi energy. Qualitatively, the DOS at
site $A_2$ is similar as in non impurity case, Fig.
\ref{FIG:u=0LDOSsingleandbilayer}(c), but it changes at site
$B_2$. There is small oscillation in the DOS close to Fermi energy
at site $B_2$.} \label{FIG:Uatav=0}
\end{figure}

\subsection{Impurity states}

We now discuss a single impurity in the absence of the external
bias. First we put an impurity at $A_1$ and study the DOS for
various values of $t_\bot$. The result of the actual calculation
is shown in Fig. \ref{FIG:Uatav=0}. The impurity has the following
effects: The DOS at the impurity site, $A_1$, decreases sharply.
The DOS at site $B_1$ increases sharply which is the signature of
a virtually bound impurity resonance state. The position of the
resonance peak shifts slightly below the Fermi energy with the
increase in $t_\perp$. This shift of the resonance peak is
expected because the $t_\perp$ also affects the pole of the Greens
function as determined by the band structure. The width of the
resonance peak decreases with $t_\bot$. This is expected because
increase in $t_\bot$ depresses the band DOS on site $A_1$, which
suppresses the scattering rate from the generated impurity state.
It corresponds to an increase of lifetime, signifying a sharp
resonance.

\begin {figure}
\includegraphics[width= 8.5 cm, height = 6.50cm]{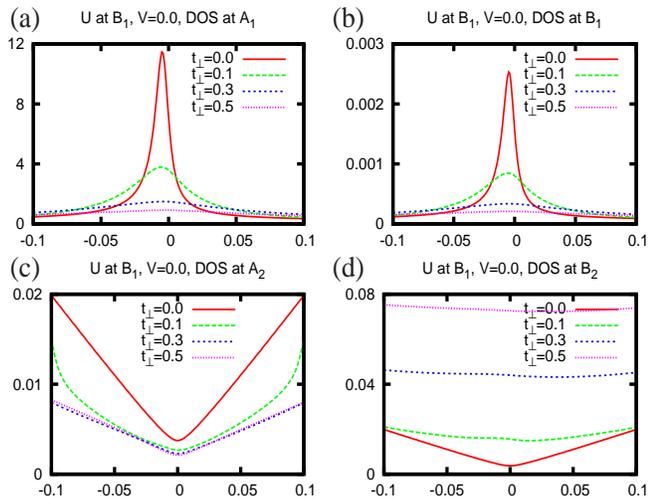}
\caption{(Color online) For single impurity at site $B_1$, DOS  as
a function of energy is shown. DOS at the impurity site, $B_1$, is
very small. DOS at the neighboring site in the same layer, $A_1$,
is very high which is the signature of the impurity resonance. The
resonance is close to the Fermi energy. Qualitatively, the DOS at
site $A_2$ is similar as in non impurity case, Fig.
\ref{FIG:u=0LDOSsingleandbilayer}c, but it changes at site $B_2$.
There is small oscillation in the DOS close to Fermi energy at
site $B_2$.}
 \label{FIG:Uatbv=0}
\end{figure}

The effect on the DOS at $A_2$ due to the impurity at $A_1$ is
very small. The overall behavior of the DOS curve does not change
as shown in Fig. \ref{FIG:Uatav=0}(c). We see a finite DOS at
$A_2$ at the Fermi energy which is nothing but the finite size
effect, which is not seen in Fig.
\ref{FIG:u=0LDOSsingleandbilayer}(c) because of the extended scale
used to show the DOS. When we zoom in and compare Fig.
\ref{FIG:u=0LDOSsingleandbilayer}(c) and Fig. \ref{FIG:Uatav=0}(c)
we see small enhancement in DOS at site $A_2$ in the presence of
impurity at $A_1$ (except for $t_\bot=0$). We see some new feature
in the DOS at $B_2$, namely the dip-hump structure in DOS close to
the Fermi energy. We believe that the small change in the DOS at
$A_2, B_2$ is due to the Friedel oscillation in the bottom layer
created due to the impurity at the top layer. The strength of the
Friedel oscillation in the bottom layer is very small because of
the geometry effect.

Now we turn to discussion on the DOS when the impurity is at site
$B_1$, Fig. \ref{FIG:Uatbv=0}. We find that the DOS at site $B_1$
is sharply reduced.  At all other sites the DOS increases compared
to the clean case, which is different from the result of the
single impurity at $A_1$ case. The DOS at $A_1$ increases sharply
which signifies the virtual bound state due to the impurity
resonance. We notice that the height of the resonance peak at site
$A_1$ decreases with the increase in $t_\perp$. Simultaneously,
the resonance becomes more broader. This effect is just opposite
to the effect seen in the case when the impurity is at site $A_1$.
This effect can be understood in terms of the band structure
corresponding to the site $B$. The increase in $t_\bot$ enhances
the band DOS on site $B_1$, which increases the scattering rate
from the generated impurity state. It corresponds to a decrease of
lifetime, signifying a broader resonance. There is no much change
in the overall behavior of the DOS at site $A_2$ compared to the
no impurity case, Fig. \ref{FIG:u=0LDOSsingleandbilayer}(c). At
site $B_2$ we see some new features, namely the dip-hump
structure. Compared to the similar effect seen at $B_2$ due to
impurity at site $A_1$, the structure of DOS in the present case
is weaker. It is because the effect gets communicated to site
$B_2$ through $A_1$ and $A_2$ which will be weaker than the
previous case where the effect was communicated through only one
bonding length.

\begin {figure}
\includegraphics[width= 8.5 cm, height = 6.50cm]{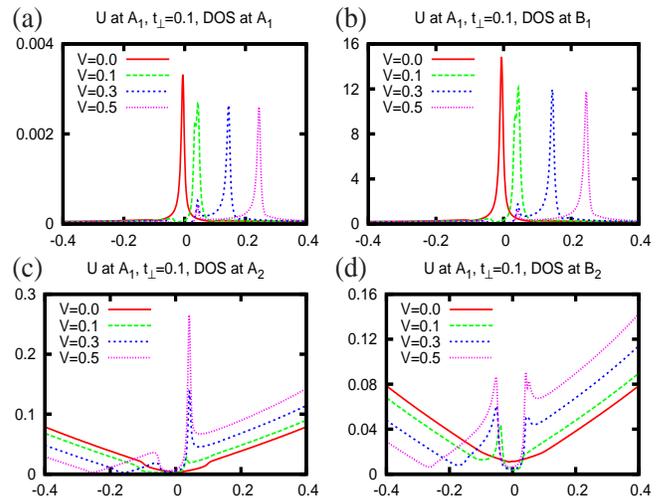}
\caption{(Color online) LDOS for single impurity at site $A_1$
with $t_\bot =0.1$ fixed for different values of $V$. There is
impurity resonance at site $B_1$. The impurity resonance is at
$E=V/2$. We also see a enhancement of the DOS at site $A_2$
assisted by the presence of impurity at $A_1$. We also see
insignificant change in DOS at site $B_2$.} \label{FIG:Uatawithv}
\end{figure}

Having discussed the effect in the single impurity case, it is
much easier to understand the double impurity case. When there are
two impurities at sites $A_1$ and $A_2$, the impurity resonance
occurs at site $B_1$ and $B_2$. The DOS at $B_1$ and $B_2$ are
identical. The height of the resonance peak increases with the
increase in $t_\perp$ and the resonance becomes sharper and
sharper. When there are two impurities at sites $B_1$ and $B_2$,
the impurity resonance occurs at site $A_1$ and $A_2$. The DOS at
$A_1$ and $A_2$ are identical. The height of the resonance peak
decreases with the increase in $t_\perp$ and the peak width
becomes broader.

Now we proceed to discuss the effect of bias on the impurity
states in bilayer graphene. When there is a single impurity at
site $A_1$, we see impurity resonance at site $B_1$, Fig.
\ref{FIG:Uatawithv}(b). The position of the resonance moves above
the Fermi energy by $V/2$. The height of the resonance decreases
slightly with the increase in the $V$. We also see small satellite
peak near the Fermi energy. At site $A_2$ also we see impurity
assisted enhancement in DOS on the gap edge, (compare with Fig.
\ref{FIG:U=0difftdiffv}c). At site $B_2$ the DOS does not show
much of qualitative change.
\begin {figure}
\includegraphics[width= 8.5 cm, height = 6.50cm]{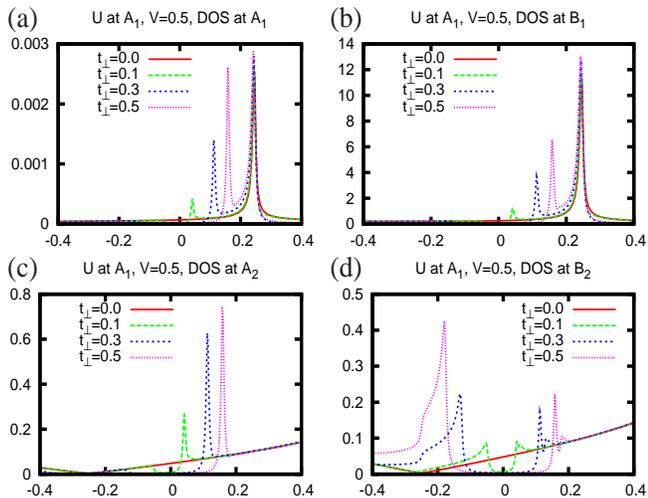}
\caption{(Color online) LDOS for a single impurity at site $A_1$,
for $V=0.5$ fixed, and with the variation in $t_\bot$.}
\label{FIG:Uatawithv=0.5fixed}
\end{figure}

To better understand the results, especially the satellite peak of
Fig. \ref{FIG:Uatawithv}(b) and the position of the enhancement
peak of Fig. \ref{FIG:Uatawithv}(c) and (d), we study the change
in LDOS due to change in $t_\bot$ for fixed $V=0.5$. The result is
shown in Fig. \ref{FIG:Uatawithv=0.5fixed}. We find some new
information from this calculation. The strength of the satellite
peak, seen in DOS at $B_1$, grows with $t_\bot$. We find that the
position of the peak is determined by both $V$ and $t_\bot$. The
calculation shows that the position of the satellite peak is at
$E_g=\frac{V}{2}\biggr{[}\frac{t_\bot^2}{t_\bot^2+V^2}\biggr{]}^2$.
This is equal to the energy gap of the second dip in the
dispersion relation of the bilayer graphene in the presence of the
external bias. The edge of the band gap seen in Fig.
\ref{FIG:U=0difftdiffv} is also at $E=E_g$. Similar behavior is
seen at site $A_2$. The enhancement of the DOS seen in Fig.
\ref{FIG:Uatawithv}(c) for $V=0.5$ increases with $t_\bot$. The
position of the peak also moves away from the Fermi energy by
$E_g$. At site $B_2$ new features are seen. The height of the
small satellite peak increases with $t_\bot$. The position of the
peak also shifts symmetrically above and below the Fermi level by
$E_g$. With the increase in $t_\bot$ a clear sign of the gap
opening is seen around the Fermi energy.

\begin{figure}
\includegraphics[width= 8.5 cm, height = 6.50cm]{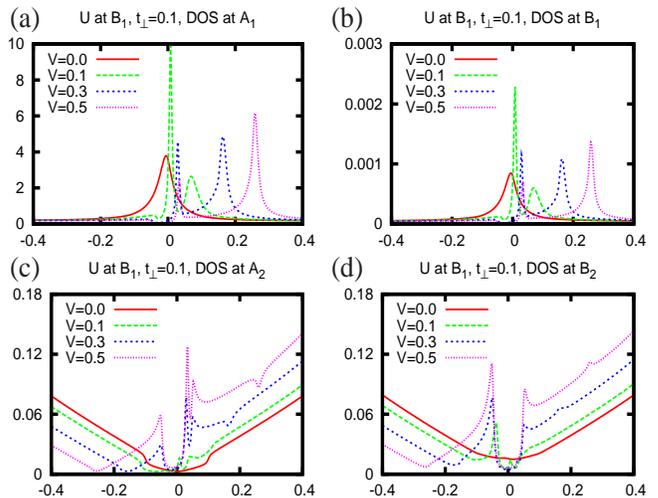}
\caption{(Color online) In addition to a resonance which follows
$V/2$, we see some interesting features at sites $A_2$, $B_2$. }
\label{FIG:Uatbwithv}
\end{figure}

When there is a single impurity at site $B_1$ (see Fig.
\ref{FIG:Uatbwithv}), we see different features than when there is
an impurity at $A_1$. At $V=0$ there is a single impurity
resonance close to the Fermi energy. When there is an external
bias, the resonance peak at site $A_1$ splits in two peaks. One of
the peak lies at $E=V/2$ above the Fermi energy. Other remains
close to the Fermi energy. With the increase in $V$, the intensity
of the peak at $V/2$ increases and that close to the Fermi energy
decreases. At site $A_2$ and $B_2$ we see impurity assisted
enhanced DOS close to the Fermi energy.

\begin{figure}
\includegraphics[width= 8.5 cm, height = 6.50cm]{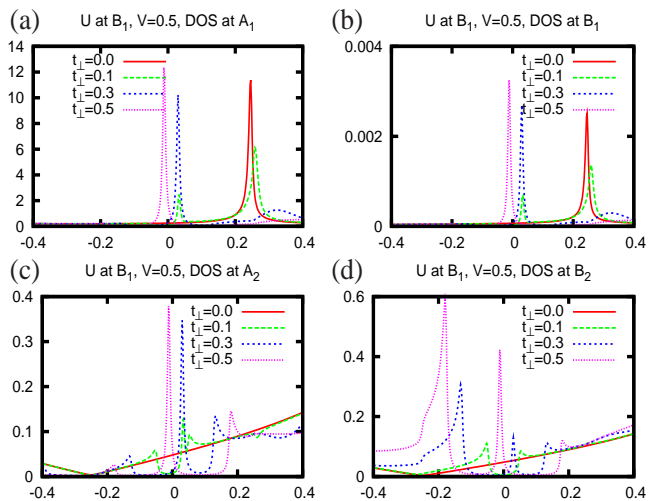}
\caption{(Color online) LDOS for a single impurity at site $B_1$,
for $V=0.5$. and with the variation in $t_\bot$. }
\label{FIG:Uatbwithv=0.5fixed}
\end{figure}

To better understand the result presented in Fig.
\ref{FIG:Uatbwithv} we perform another calculation for LDOS at
fixed $V=0.5$ and different $t_\bot$. The result is shown in Fig.
\ref{FIG:Uatbwithv=0.5fixed}. In the Fig.
\ref{FIG:Uatbwithv=0.5fixed}(a) we can see that the weight of LDOS
at $E=V/2$ decreases with $t_\bot$. This loss of weight is
transferred close to the Fermi energy. This phenomenon can be
understood in terms of the change in LDOS (at fixed $V=0.5$ in the
absence of the impurity) due to the change in $t_\bot$, (see Fig.
\ref{FIG:U=0difftdiffv}(e), and \ref{FIG:U=0difftdiffv}f)), where
we see that finite $t_\bot$ opens up a gap around the Fermi energy
and the gap becomes more and more well defined for increased
$t_\bot$. This gap leads to a new resonance state around the Fermi
energy in the presence of the impurity. The shift of the weight of
the LDOS close to the Fermi energy at site $A_2$ and $B_2$, can
also be understood using the same logic. We can also see in Fig.
\ref{FIG:U=0difftdiffv}(f) that the LDOS at site $B_1$ at $E=V/2$
increases with the increase in $t_\bot$ which explains the
broadening effect seen in Fig. \ref{FIG:Uatbwithv=0.5fixed}(a)
near $E=V/2$.

When we have two impurities at $A_1$ and $A_2$ we see impurity
resonances at $B_1$  and $B_2$. The position of the impurity
resonance is again determined by $V/2$. Interestingly, the
impurity resonance at site $B_1$ lies above the Fermi energy
whereas that at site $B_2$ is below the Fermi energy. The DOS at
$B_1$ can be obtained by mirror reflection of the DOS at $B_2$
about the axis of $E=0$.

When we put two impurities at $B_1$ and $B_2$ impurity resonance
arises at $A_1$  and $A_2$. The position of the impurity resonance
is again determined by $V/2$. The impurity resonance at site $A_1$
lies above the Fermi energy whereas that at site $A_2$ is below
the Fermi energy. The DOS at $A_1$ can be obtained by mirror
reflection of the DOS at $A_2$ about the axis of $E=0$.

\section{Concluding remarks}

Before concluding, we would like to make two comments. To our
knowledge two studies have been done which resembles to our work.
Wang et. al.\cite{wang2007} have studied the effects of voids in
the absence of external bias. In particular they investigated the
quantum interference pattern in a specific energy and no impurity
induced resonance states were discussed. Similarly, the disorder
problem in biased bilayer graphene was studied by Nilsson and
Castro Neto.\cite{nilsson2007a} In our work we have addressed the
issue about the sensitivity of the impurity induced resonance
states to the underlying electronic band structure in a bilayer
graphene, through the application of pressure or/and electric
bias, which were not touched in these literatures.

In conclusion we study the possibility of tuning impurity states
in a bilayer graphene. We systematically study the signature of
the impurity resonance states by looking at the local density of
states at four inequivalent sites of the bilayer graphene. This
work should be the most detailed study of the impurity states in
bilayer graphene. We have shown that in bilayer gaphene the
impurity states can be tuned using an external bias and changing
the interlayer hopping energy. Our predictions about the evolution
of the impurity states in bilayer graphene can be tested by the
scanning tunnelling microscopy.

Acknowledgements: One of us (H.D.) is grateful to K. Bedell and
the Boston College for financial support. This work was supported
by DOE at Los Alamos under Grant No. DE-AC52-06NA25396 and the
LDRD programs.

\bibliographystyle{apsrev}
\bibliography{references}
\end{document}